\documentclass{JHEP}
\usepackage{graphicx}		 
\newcommand{\be}{\begin{equation}}
\newcommand{\ee}{\end{equation}}
\newcommand{\ba}{\begin{eqnarray}}
\newcommand{\ea}{\end{eqnarray}}

\title{Cosmology: theory\footnote{Plenary talk delivered at the
 European Physical Society  Conference on High Energy Physics, 18-24
 July, 2013, Stockholm, Sweden}}

\author{Mikhail SHAPOSHNIKOV\\ 
        Institut de Th\'eorie des Ph\'enom\`enes Physiques\\
        \'Ecole Polytechnique F\'ed\'erale de Lausanne\\
	CH-1015 Lausanne\\
	Switzerland\\
        E-mail: \email{mikhail.shaposhnikov@epfl.ch}}

\abstract{The discovery of $126$ GeV Higgs boson and observations of
no signs of new physics at the LHC  implies that the Standard Model
of elementary particles is a self-consistent weakly-coupled effective
field theory all the way up to the Planck scale without the addition
of any new particles. I will discuss possible consequences of these
findings for cosmology.}

\begin{document}

\section{Introduction}
\label{intro}

It is impossible to cover the wast field of cosmological theory in 30
minutes. So, a number of interesting topics, such as  dark energy,
quintessence,  non-Gaussian inflationary perturbations, cosmological
magnetic fields, primordial nucleosynthesis, string cosmology, quantum
gravity, massive gravity, galileons, chameleons, radion cosmology,
modified gravity, $f(R)$ gravity, Ho\u{r}ava-Lifshitz gravity,
landscape and multiverse, holographic cosmology and a number of
others, will not be addressed.

I will discuss here several cosmological issues that are related to a
recent discovery of the $126$ GeV Higgs boson at the LHC and to
non-observation of signs of new physics at the LHC or elsewhere.

The paper is organised as follows. In Section \ref{hmass} we discuss 
what the findings at the LHC may mean for cosmology in general, in
Section \ref{infl} we consider cosmological inflation and the
possibility that it may be associated with the Higgs boson of the
Standard Model. In Section \ref{baudm} we discuss  baryon asymmetry of
the Universe (BAU)  and Dark Matter (DM) in the view of the LHC
results. In Section \ref{dr} we overview cosmological constraints on 
neutrino masses and  dark radiation. Section \ref{concl} is
conclusions.

\section{LHC, Higgs mass and Cosmology}
\label{hmass}

\begin{figure}[t]
\vspace*{-1.3cm}
\includegraphics[width=0.45\textwidth]
{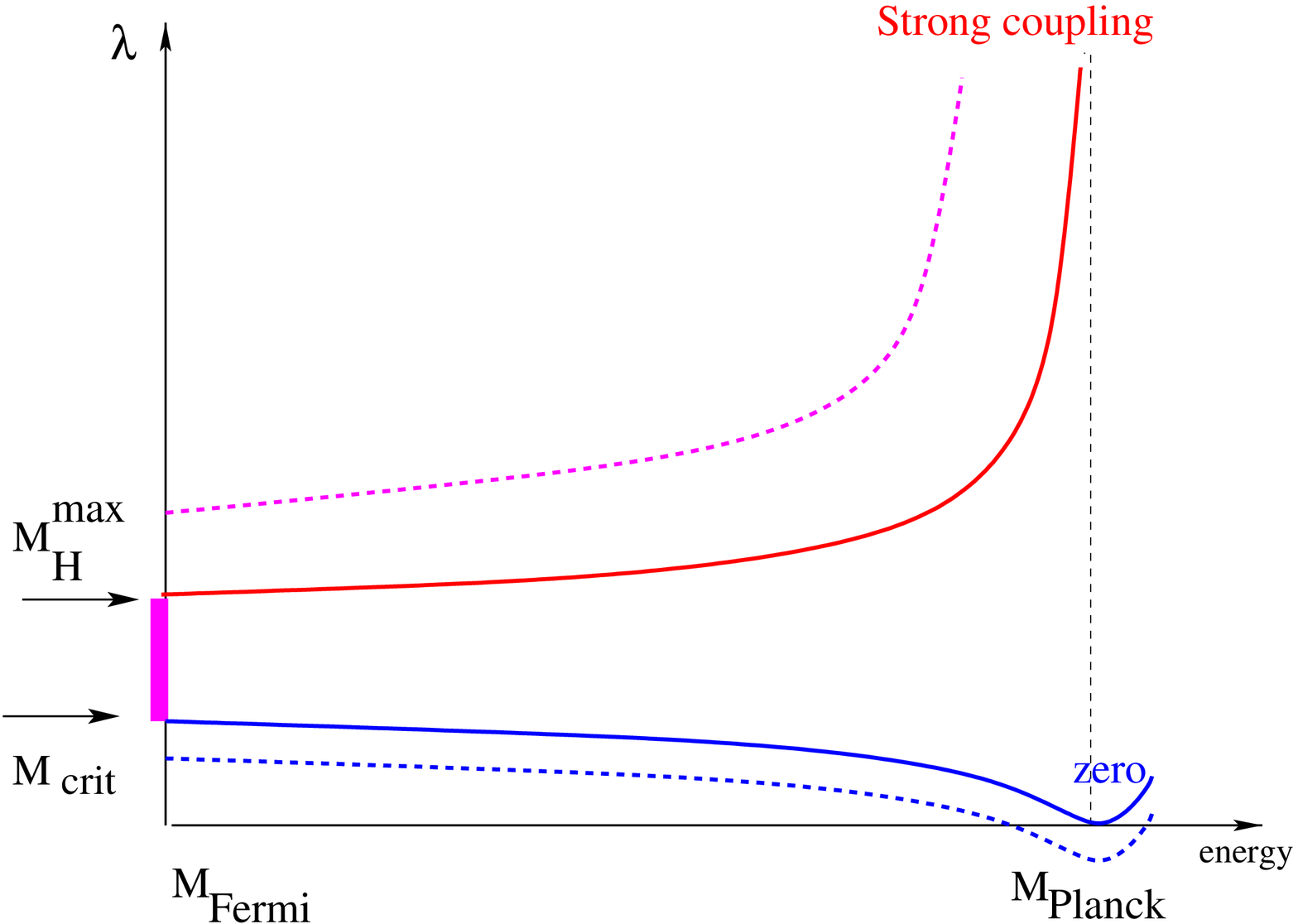}\hspace{0.7cm}
\raisebox{1.2cm}{\includegraphics[width=0.35\textwidth]
{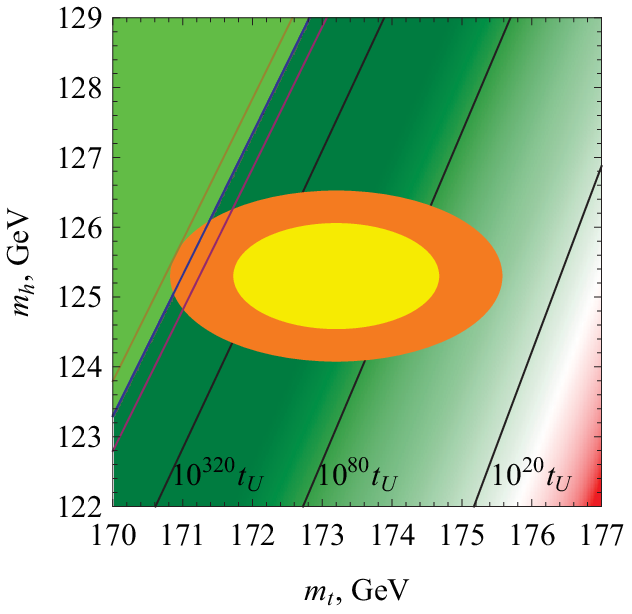}}
\caption{
Left panel: Different patters of the behaviour of the Higgs self
coupling with energy. For $M_H> M_H^{\rm max}$ the Landau pole appears
at energies below the Planck scale. If $M_H < M_{\rm crit}$ the scalar
constant becomes negative at energies below the Planck mass, and
electroweak vacuum becomes metastable. Right panel (courtesy of F.
Bezrukov): The lifetime of the electroweak vacuum as a function of top
quark and Higgs boson masses. Ellipses correspond to 1 and 2 $\sigma$
contours in $M_H$ and $m_t$, $t_U$ is the age of the Universe. Along
the straight lines the lifetime of the vacuum is given by the number
in the plot. The light green region in the upper left  corner
corresponds to the stable vacuum. 
}
\label{vacuum}
\end{figure}

After the discovery of the Higgs boson  at the LHC by ATLAS
\cite{Aad:2012tfa} and CMS \cite{Chatrchyan:2012}, the last missing
particle of the Standard Model (SM) has been found. At present, the
main LHC result can be formulated as follows: the SM is a
self-consistent, weakly coupled effective field theory all the way up
to the Planck scale. First, no significant deviations from the SM
predictions are seen and no convincing signal in favour of existence
of  new physics beyond the SM is observed. Second, the mass of the
Higgs boson $M_H$ is smaller than $M_H^{\rm max} = 175$ GeV. If this
were not the case, the Landau pole in the Higgs scalar self-coupling
would be below the Planck quantum gravity scale $M_P = 2.44\times
10^{18}$ GeV (see, e.g. \cite{Ellis:2009tp}), calling for an extension
of the SM at some energies between Fermi and Planck scales. Finally, 
the mass of the Higgs is sufficiently large, $M_H > 111$ GeV, meaning
that  our  vacuum is stable or metastable with a lifetime greatly
exceeding the Universe age \cite{Espinosa:2007qp}. The behaviour of
the Higgs boson self-coupling $\lambda$  as the function of  energy
and the life-time of the Universe as a function of the Higgs boson and
top quark masses are shown in   Fig. \ref{vacuum}.

\begin{figure}[t]
\centerline{\includegraphics[width=0.8\textwidth]{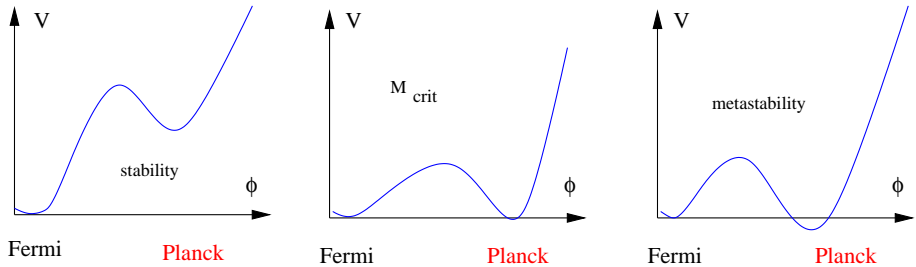}}
\caption{
The form of the effective potential for the Higgs field $\phi$ which
corresponds to the stable (left), critical (middle) and metastable
(right) electroweak vacuum. The form of the effective potential is
tightly related to the energy dependence of the Higgs self-coupling
constant $\lambda(\mu)$: the potential is negative almost in the same  
domain where $\lambda(\phi)< 0$. 
}
\label{stab}
\end{figure}

The mass of the Higgs boson, found experimentally, 
($M_H=125.5\pm0.2_{stat}~^{+0.5}_{-0.6}$$_{syst}$ GeV, $\rm ATLAS$
\cite{Aad:2012tfa} $M_H=125.7\pm 0.3_{stat}\pm0.3_{syst}$ GeV, CMS
\cite{Chatrchyan:2012}) is very close to the ``critical Higgs mass''
$M_{\rm crit}$, which appeared in the literature well before the
Higgs discovery in different contexts. The value of $M_{\rm crit}$ is
the stability bound on the Higgs mass $M_H>M_{\rm crit}$, see Fig.
\ref{stab} (the ``multiple point principle'', put forward in
\cite{Froggatt:1995rt}, leads to prediction $M_H=M_{\rm crit}$), to the
lower bound on the Higgs mass coming from requirement of the Higgs
inflation \cite{Bezrukov:2009db,DeSimone:2008ei}, and to the
prediction of the Higgs mass coming from asymptotic safety scenario
for the SM \cite{Shaposhnikov:2009pv}. 

The computation of numerical value of $M_{\rm crit}$ was a subject of
many papers, the most recent result is convenient to write in the
form\footnote{Note that this form is different from the original
works, as well as the uniform estimates of the theoretical errors,
which are the sole responsibility of the present author.} 
\be
M_{\rm crit}= [129.3 + \frac{y_t(m_t) -
0.9361}{0.0058}\times 2.0 -\frac{\alpha_s(M_Z)-0.1184}{0.0007}\times
0.5]~{\rm GeV}~. 
\label{mcrit}
\ee 
Here $y_t(m_t)$ is the top Yukawa coupling in $\rm{\overline{MS}}$
renormalisation scheme taken at the pole mass of top quark $m_t$, and
$\alpha_s(M_Z)$ is the QCD coupling at the $Z$-boson mass. The
computation consists of matching of $\rm{\overline{MS}}$ parameters of
the SM to the physical parameters such as the masses of different
particles  and then renormalisation group running of coupling
constants to high energy scale, supplemented by the computation of the
effective potential for the Higgs field. All recent works
\cite{Bezrukov:2012sa,Degrassi:2012ry,Buttazzo:2013uya} used 3-loop
running of the coupling constants found in
\cite{Mihaila:2012fm}-\cite{Bednyakov:2013eba};
ref. \cite{Bezrukov:2012sa} accounted  for ${\cal O}(\alpha\alpha_s)$
corrections to the matching procedure, getting  $129.4$ GeV for the
central value of $M_{\rm crit}$ with the theoretical error  $1.0$ GeV,
ref. \cite{Degrassi:2012ry} got $129.6$ GeV with smaller error  $0.7$
GeV, accounting for ${\cal
O}(\alpha\alpha_s,y_t^2\alpha_s,\lambda^2,\lambda\alpha_s)$ terms in
the matching, while the complete analysis of 2-loop corrections in
\cite{Buttazzo:2013uya} gives $129.3$ GeV for the central value with
very small theoretical error $0.07$ GeV.

\begin{figure}[t]
\vspace*{-2cm}
\centerline{
\raisebox{2cm}{\includegraphics[width=0.5\textwidth]{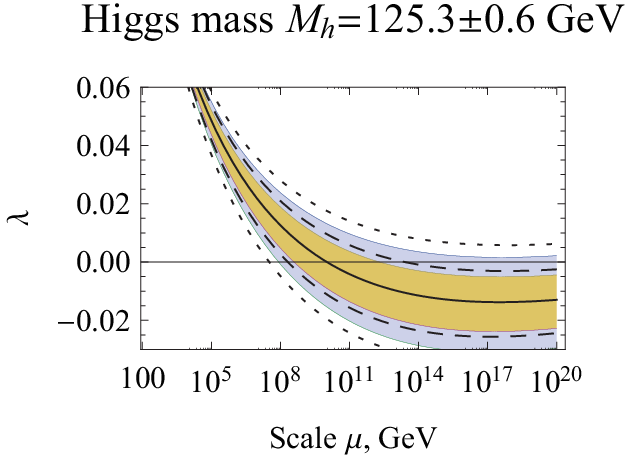}}\hspace{0.4cm}
\raisebox{1.5cm}{\includegraphics[width=0.4\textwidth]{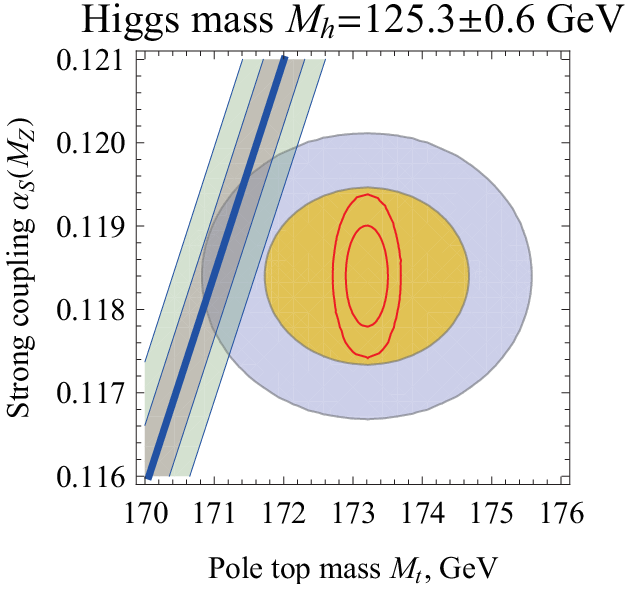}}
}
\caption{
Both panels: The  shaded regions account for $1$ and $2$ $\sigma$ 
experimental uncertainties in $\alpha_s$ and the pole top quark mass
$m_t$, and theoretical errors in extraction of $y_t$ from experiment.
Left panel: Running of the scalar self coupling constant with energy.
Dashed and dotted lines correspond to varying in addition the mass of
the Higgs boson within 1 and 2 $\sigma$ experimental errors. Right
panel:  The blue line gives the relation between $\alpha_s$ and the
pole top mass following from eq. (2.1) if $M_H$ is identified with
$M_{\rm crit}$. The shaded regions around it correspond to 1 and 2
$\sigma$ experimental errors in the Higgs mass. Red ellipses
correspond to the accuracy achievable at $e^+e^-$ collider
\cite{Alekhin:2012py}. 
}
\label{lambda}
\end{figure}

At present, {\em we do not know} whether our vacuum is stable or
metastable. Fig. \ref{lambda} shows the behaviour of the scalar
self-coupling within experimental and theoretical uncertainties,
together with confronting the value of $M_{\rm crit}$ from eq.
(\ref{mcrit}) with the data. For making these plots, the pole top mass
was taken from the Tevatron \cite{CDF:2013jga},
$m_t=173.2\pm0.51_{stat}\pm 0.71_{sys}$ GeV (the combined ATLAS and
CMS value is  $m_t=173.4\pm0.4\pm 0.9$ GeV \cite{Battilana:2013aia}),
and the value of  $\alpha_s(M_Z)=0.1184\pm0.0007$
\cite{Beringer:1900zz}.

To determine the relation between $M_{\rm crit}$ and $M_H$, the
precision measurements of  $m_H, y_t$ and $\alpha_s$ are needed. The
main uncertainty is in the value of top Yukawa coupling, $y_t$. In
general, an $x$ GeV experimental error in  $m_t$ leads to  $\simeq
2\times x$
GeV error in  $M_{\rm crit}$. The difficulties in extraction of $y_t$ from
experiments at the LHC or Tevatron are discussed in
\cite{Alekhin:2012py}. Here we just mention that the  non-perturbative
QCD effects, $\delta m_t\simeq \pm \Lambda_{QCD} \simeq \pm 300$ MeV
lead to $\delta M_{\rm crit}\simeq  \pm0.6$ GeV. The similar in amplitude
effect comes from (unknown) ${\cal O}(\alpha_s^4)$ corrections to the
relation between the pole and $\rm{\overline{MS}}$ top quark masses.
According to \cite{Kataev:2010zh}, this correction can be as large as 
$\delta y_t/y_t \simeq -750 (\alpha_s/\pi)^4\simeq - 0.0015$, leading
to  $\delta M_{\rm crit}\simeq - 0.5$ GeV.

What do the (meta) stability of our vacuum and the agreement
of the Standard model with the LHC experiments mean for cosmology?  We
can consider  two different possibilities.\\
 (i) The Higgs mass is smaller than $M_{\rm crit}$, so that the 
scalar self coupling crosses zero at energy scale  $M_\lambda \ll
M_P$, where  $M_\lambda$ can be as ``small'' as  $10^8$ GeV, within
the experimental and theoretical error-bars, see Fig. \ref{lambda}. \\
 (ii) The Higgs mass is larger or equal to  $M_{\rm crit}$, and  the
Higgs self coupling never crosses zero (or does so close to the Planck
scale, where gravity effects must be taken into account),  see Fig.
\ref{lambda}.

If (i) is realised, there are two ways to deal with the metastability
of our vacuum. The first one is cosmological: it is sufficient that 
the Universe after inflation finds itself in our vacuum with 
reheating temperature below $M_\lambda$. Then this guarantees that we
will stay in it for a very long time. This happens, for example, in
$R^2$ inflation \cite{Bezrukov:2011gp}. The another possibility is
related to possible existence of new physics at $M_\lambda$ scale,
which makes our vacuum unique (see, e.g. \cite{EliasMiro:2012ay}). 

If (ii) is realised, then  no new physics is needed between the Fermi
and Planck scales. 

In the rest of this talk I will concentrate on the second, most
conservative option and assume that there is no new physics between
the Fermi  and  Planck scales. If it is realised, we have to address a
number of questions anew: ``What drives inflation?'', ``What is the origin
of baryon asymmetry of the Universe?'', ``What is the dark matter
particle?''

\section{Inflation}
\label{infl}
It is well known that for inflation we better have some bosonic field,
which drives it (for a review see, e.g. \cite{Linde:2007fr}). At last,
the Higgs boson has been discovered. Can it make the Universe flat,
homogeneous, and isotropic, and produce the necessary spectrum of
fluctuations for structure formation? The answer to this question is
affirmative \cite{Bezrukov:2007ep}.

The main idea of Higgs inflation is related to a non-minimal coupling of
the Higgs field to gravity, described by the action 
\be 
   S_G =\int d^4x \sqrt{-g} \Bigg\{-\frac{M_P^2}{2}R
  - \frac{\xi |\phi|^2}{2}R
  \Bigg\}~.
\ee
Here $R$ is the scalar curvature, the first term is the standard
Hilbert-Einstein action, $\phi$ is the Higgs field, and $\xi$ is a new
coupling constant, fixing the strength of ``non-minimal'' interaction.
This constant cannot be fixed by a theoretical computation, but its
presence is actually required for consistency of the SM in curved
space-time (see, e.g. \cite{Feynman:1996kb}). 

Consider now large Higgs fields, typical for chaotic inflation
\cite{Linde:1983gd}. Then the gravity strength, given by the effective
Planck mass in the Higgs background, is changed as  $M_P^{\rm eff} =
\sqrt{M_P^2+ \xi |\phi|^2} \propto |\phi|$. In addition, all particle
masses are also proportional to the Higgs field. This means that  for
$|\phi|\gg\frac{M_P}{\sqrt{\xi}}$ the physics does not depend on the
value of the Higgs field, as all dimensionless ratios are $|\phi|$
independent. This leads to an existence of the flat direction for a
canonically normalized scalar field $\chi$, related to the Higgs field
by conformal transformation. After inflation with $N \simeq 58$
e-foldings the energy of the Higgs field is transferred to other
particles of the SM, reheating the Universe up to the temperature 
$T_{reh}\sim 10^{13-14}$ GeV  \cite{Bezrukov:2008ut,GarciaBellido:2008ab}.

For the Higgs inflation to work, the scalar self-coupling constant
$\lambda$ must be positive up to the scale of inflation $\mu_{infl} =
M_P/\sqrt{\xi}$. Numerically, this leads to the constraint $M_H>M_{\rm
crit}$ with extra theoretical uncertainty of $\delta M_H\sim 1$ GeV
\cite{Bezrukov:2010jz}. Though the theory in the electroweak vacuum
enters into strong coupling regime at energies smaller than the Planck
scale by a factor $\xi$ \cite{Barbon:2009ya,Burgess:2010zq}, the
analysis of higher dimensional operators and radiative corrections at
large Higgs background, necessary for inflation, shows that  the Higgs
inflation occurs in the weak coupling regime and is self-consistent
\cite{Bezrukov:2010jz}.

The cosmological predictions of the Higgs inflation can be compared 
with observations performed by the Planck satellite.  The
Higgs-inflaton potential depends on one unknown parameter, $\xi$. It
can be fixed by the amplitude of the CMB temperature fluctuations 
$\delta T/T$ at the WMAP normalization  scale $\sim 500$ Mpc,  with
the use of precise knowledge of the top quark and Higgs masses, and
$\alpha_s$. In general,  $\xi > 600$ \cite{Bezrukov:2009db}. Since the
Higgs mass lies  near $M_{\rm crit}$, the actual value of $\xi$ may be
close to the lower bound.

\begin{figure}[t]
\vspace*{-1cm}
\raisebox{0.7cm}{\centerline{\includegraphics[width=0.5\textwidth]{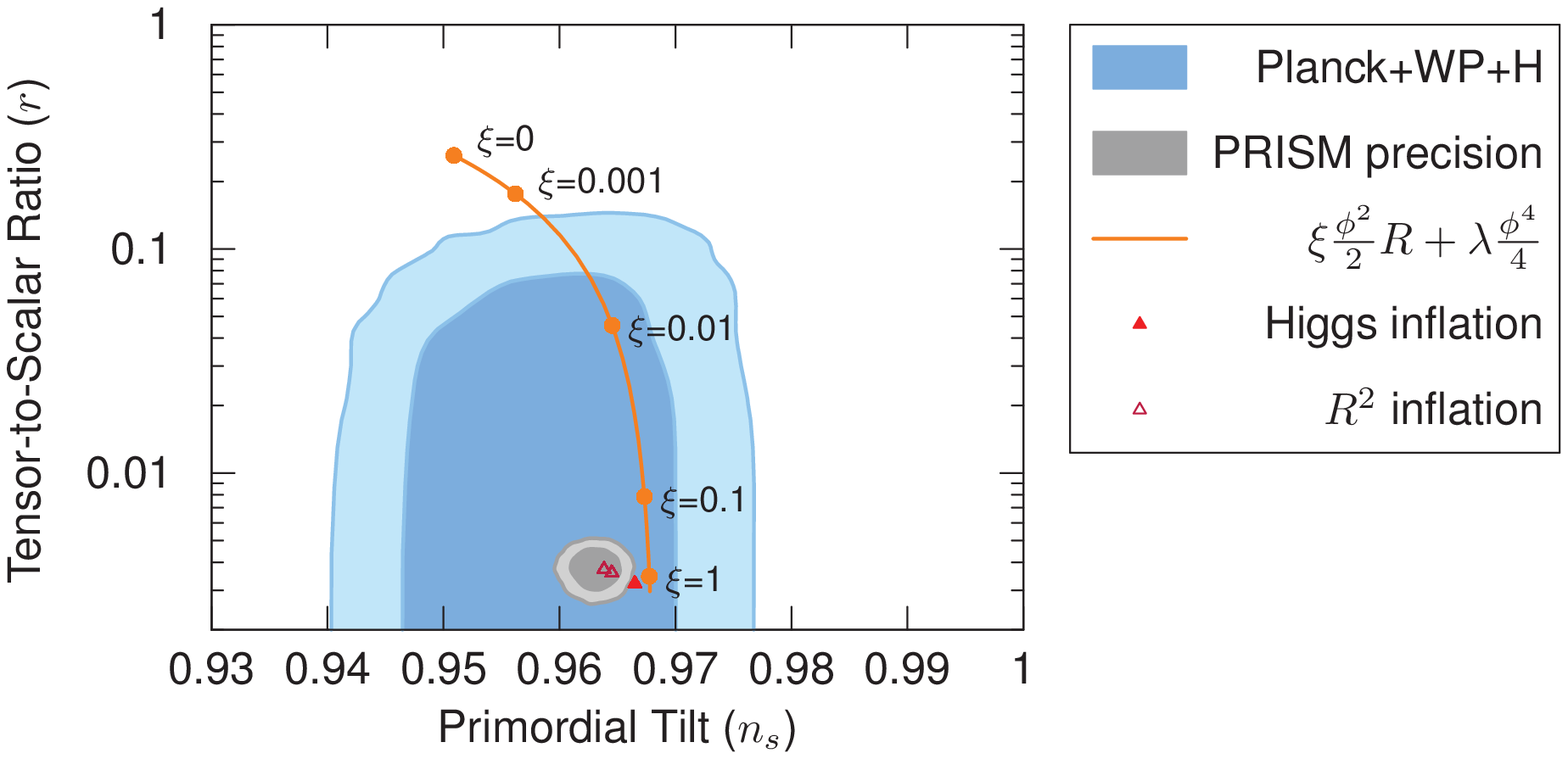}
\includegraphics[width=0.5\textwidth]{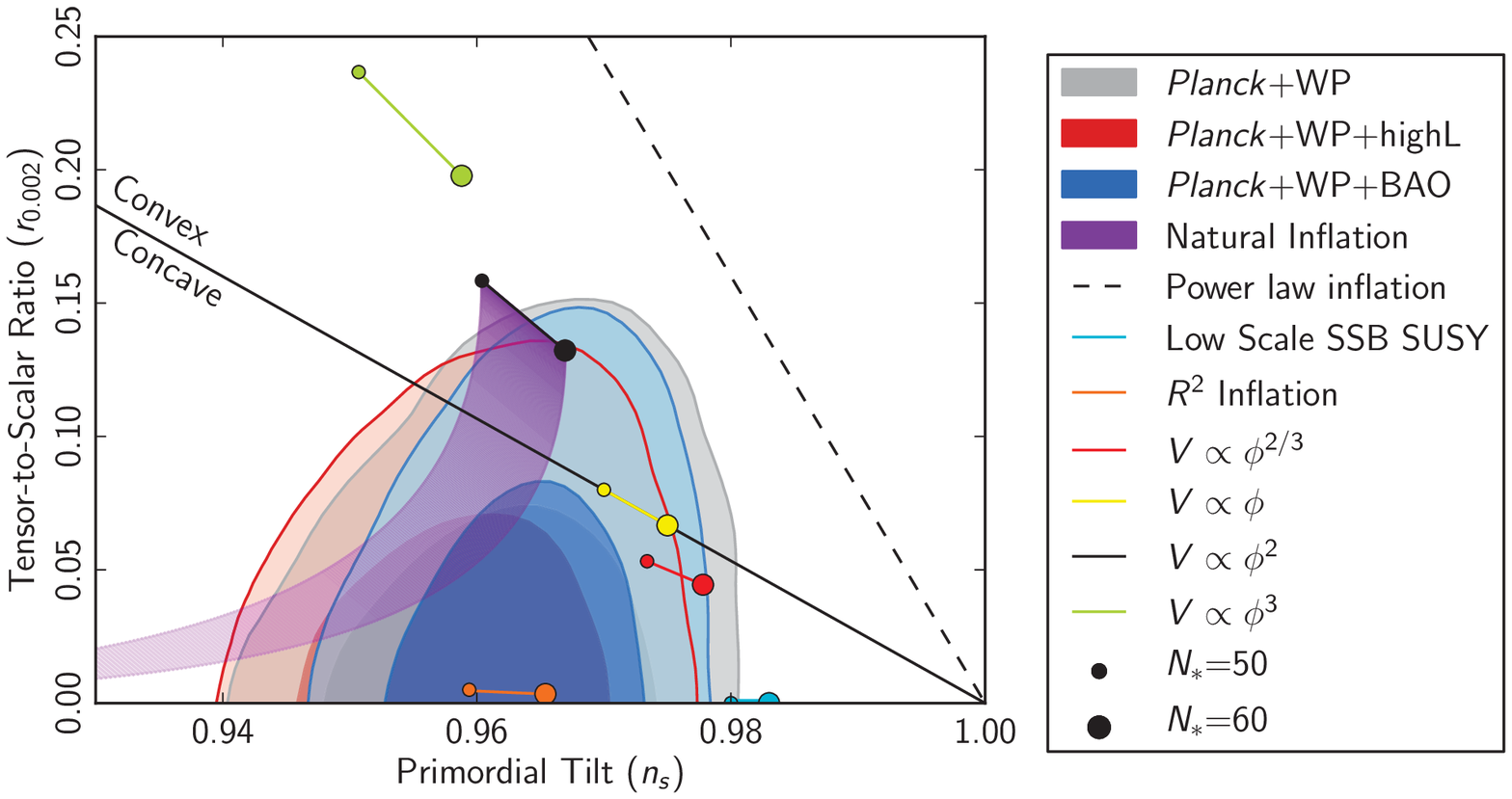}}}
\caption{Left panel (courtesy of F. Bezrukov):  The predictions of
different inflationary models, closely related to the Higgs inflation,
versus observations. The red line  corresponds to a theory with light
inflaton and small non-minimal coupling to gravity
\cite{Bezrukov:2013fca}. The predictions of $R^2$ inflation
\cite{Starobinsky:1980te,Mukhanov:1981xt} can be distinguished from the
Higgs inflation with PRISM mission. Right panel: Predictions of
different inflationary models contrasted with the Planck results (from
ref. \cite{Ade:2013uln}).
}
\label{cmb}
\end{figure}

Also,  the value of spectral index $n_s$ of scalar density perturbations
\be
\left\langle\frac{\delta T(x)}{T} \frac{\delta T(y)}{T}
\right\rangle\propto\int
\frac{d^3k}{k^3}e^{ik(x-y)}k^{n_s-1}
\ee
and the amplitude of tensor perturbations $r =
\frac{\delta\rho_s}{\delta\rho_t}$ can be determined. The predictions,
together with the Planck results,  are presented in Fig. \ref{cmb},
and are well inside the 1 sigma experimental contour. Moreover, as for
any single field inflationary model, the  perturbations are Gaussian,
in complete agreement  with Planck \cite{Ade:2013uln} (for discussion
of Planck results see Rosset talk at the parallel Session). 

There is a number of inflationary models which give predictions for
$n_s$ and $r$ similar to those of the Higgs inflation. This is true,
in particular, for the modification of the theory in the gravitational
sector by adding the $R^2$ term
\cite{Starobinsky:1980te,Mukhanov:1981xt}. The $R^2$ theory contains
an additional degree of freedom -- scalaron, with the mass $M_S\simeq
3\times 10^{13}$ GeV\footnote{This particle creates an extra
fine-tuning problem, as its loops shift the Higgs mass by $\delta
M_H^2 \sim \frac{1}{16\pi^2} \frac{M_S^4}{M_P^2} \sim 10^{14}$
GeV$^2\gg M_H^2$.}. It has somewhat smaller reheating temperature
$\sim 3\times 10^9$ GeV and smaller number of e-foldings, $N\simeq 54$
\cite{Vilenkin:1985md, Gorbunov:2010bn}. The generalisations of the
Higgs and $R^2$ inflation to the case of supergravity were discussed
in
\cite{BenDayan:2010yz}-\cite{Ketov:2012jt}.
For more extended discussion of inflationary model see talk by
Bezrukov at the parallel Session.

\section{Baryon asymmetry of the Universe, dark matter and the LHC}
\label{baudm}
One of the most popular mechanisms for for creation of the
cosmological baryon excess is electroweak baryogenesis. In short, its
idea is as follows \cite{Nelson:1991ab}. At high temperatures the
Universe is in the symmetric phase of the SM. During the Universe
cooling the first order phase transition converting the symmetric
phase to the Higgs phase occurs. The phase transition  goes through 
nucleation of bubbles of the Higgs phase. The scattering of different
particles on the domain walls leads to the spacial separation of
baryon number: an excess of baryons inside the bubbles survives till
the present time, whereas the excess of antibaryons outside the
bubbles is destroyed by  sphalerons
\cite{Kuzmin:1985mm,Klinkhamer:1984di}. 

This mechanism does not work in the SM. First, there is no phase
transition for the Higgs boson with the mass above $75$ GeV
\cite{Kajantie:1996mn}. Second, the SM CP-violation is most probably
too weak \cite{Shaposhnikov:1987tw,Gavela:1994dt,Huet:1994jb} (see
however \cite{Farrar:1993hn}). With the new LHC constraints on SUSY
the electroweak baryogenesis 
is challenged, but still possible in the Minimal Supersymmetric
Standard Model 
\cite{Curtin:2012aa,Carena:2012np,Cohen:2012zza}.

The another popular mechanism for baryogenesis is  thermal
leptogenesis \cite{Fukugita:1986hr}. Here the  superheavy Majorana
leptons with the mass  $\sim 10^{10}$ GeV decay and produce lepton
asymmetry, which is converted to baryon asymmetry by sphalerons. The 
necessity of having superheavy particles leads to the hierarchy
problem \cite{Vissani:1997ys}: their loops shift the Higgs mass
towards the Grand Unified scale. This may be cured by low energy SUSY,
but no signs of it were seen at LEP, Tevatron or LHC. A possible way
out is the resonant leptogenesis with degenerate Majorana leptons and
relatively small masses  $\sim 1$ TeV \cite{Pilaftsis:2003gt}. In any
event, the thermal leptogenesis cannot be disproved experimentally,
but it is fine tuned without new physics at the Fermi scale.

So, if the next LHC runs will continue to confirm the SM, the popular
mechanisms for baryogenesis will be disfavored. 

Now, let us come to dark matter. The most popular DM candidate is the
Weakly Interacting Massive Particle -- WIMP, associated with new
physics solving the hierarchy problem at the electroweak scale. If no
new physics will be discovered at the LHC, this candidate will not be
that attractive anymore.

Therefore, it is timely to readdress the problems of baryon asymmetry
of the Universe and of dark matter in case there is no new physics
between the Fermi and Planck scales.

\begin{figure}[t]
\vspace*{2.2cm}
\centerline{
\includegraphics[width=0.4\textwidth]{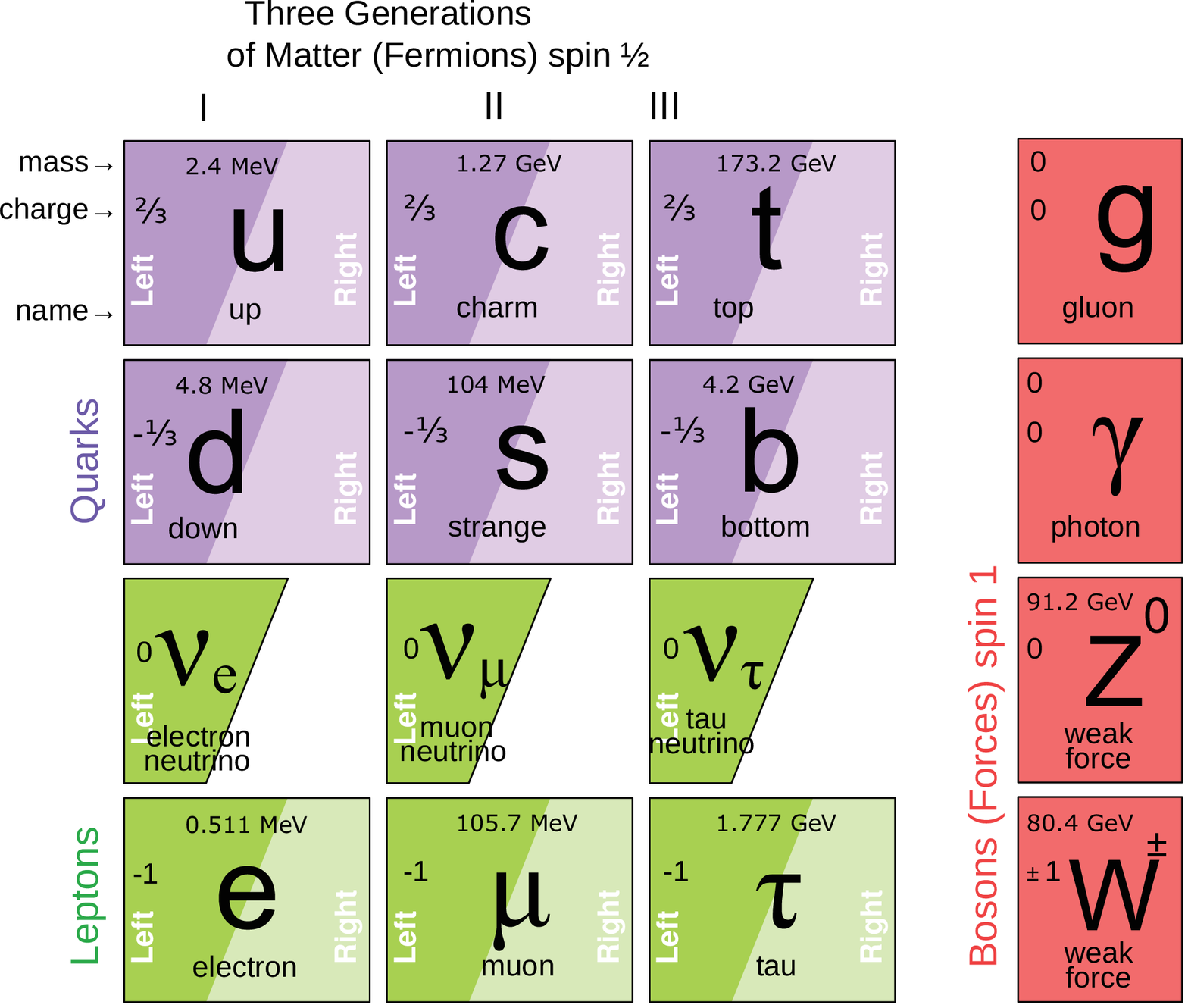}
\hspace*{1cm}\includegraphics[width=0.4\textwidth]{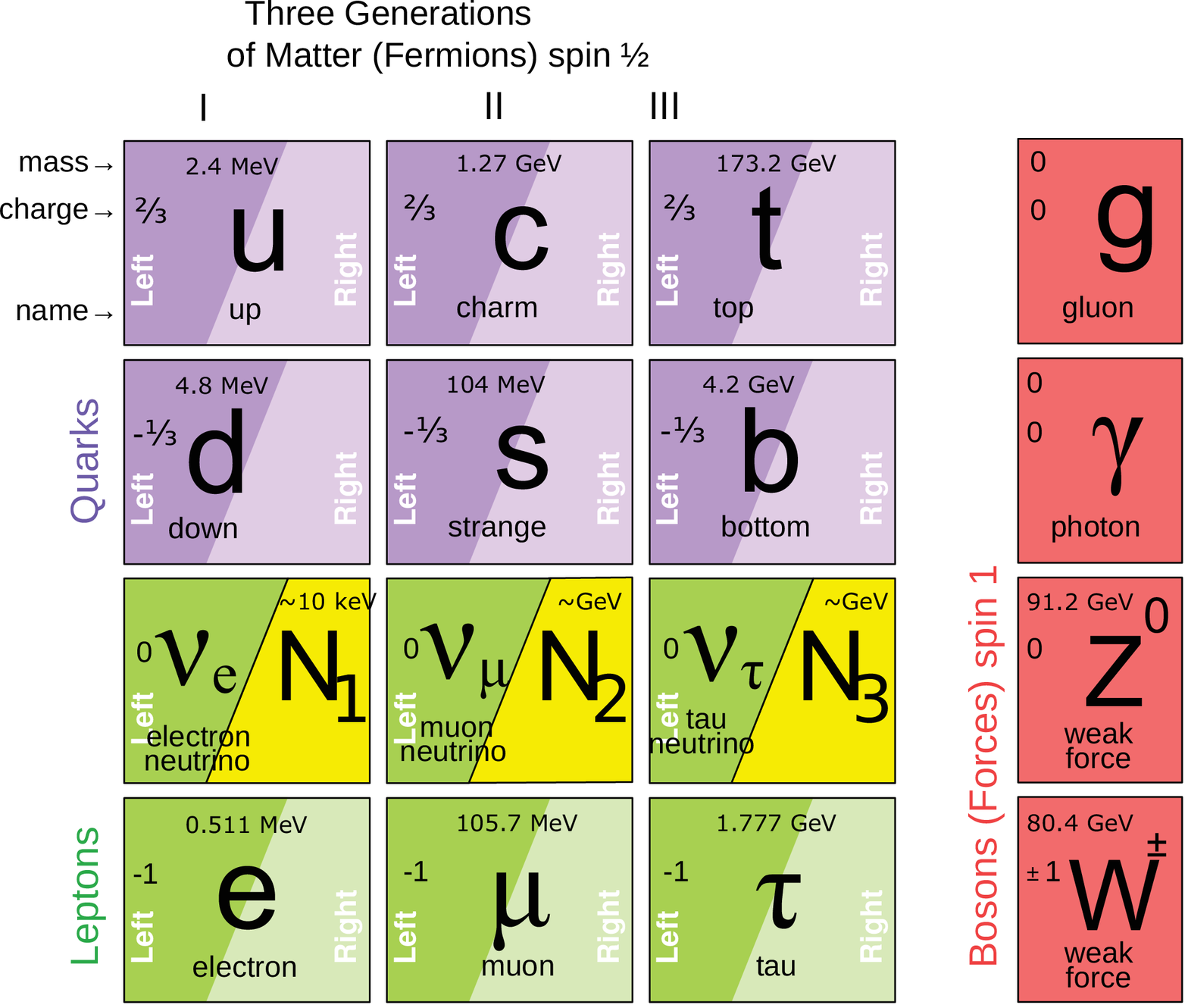}
}
\vspace*{-2.2cm}
\caption{Particle content of the SM and its minimal extension in the
neutrino sector. In the  SM the right-handed partners of
neutrinos are absent. In the  $\nu$MSM all fermions have both
left- and right-handed components and masses below the Fermi scale.}
\label{nuMSM}
\end{figure}

A possible solution to these problem is given by the Neutrino Minimal
Standard Model ($\nu$MSM) \cite{Asaka:2005an,Asaka:2005pn} (for
extensive discussion see Ruchayskiy talk at the Cosmology parallel
Session, the review \cite{Boyarsky:2009ix} and references therein). It
adds to the SM three Majorana leptons $N_I$ -- right-handed partners
of active neutrinos, see Fig. \ref{nuMSM}. The $N_1$ with the mass in
keV region  plays the role of dark matter particle, which can be
searched for  with the use of X-ray telescopes. The  role of
$N_2,~N_3$ with masses in 100 MeV -- GeV region is to  ``give'' masses
to neutrinos via electroweak scale see-saw mechanism and produce BAU.
They can be searched for at fixed target experiments with the use of
intensive proton beams, such as SPS at CERN \cite{Bonivento:2013jag}.

\section{Dark radiation, neutrino masses}
\label{dr}
There are other predictions of the Standard Model where it can be
compared with cosmological observations. There are four very light or
massless particles in the SM: 3 neutrinos and a photon. The Standard
cosmology leads to prediction of the number of relativistic degrees
of  freedom (photon is not included) in terms of ``effective number of
neutrino species'' $N_{eff}=3.046$. The deviation of this number from
$3$ is due to non-instantaneous decoupling and finite temperature
effects (for a review  see, e.g. \cite{Lesgourgues:2006nd}). Also, the
analysis of neutrino oscillations provides the lower bound on the sum
of neutrino masses (see, e.g. \cite{GonzalezGarcia:2012sz}): $\sum
m_\nu > 0.06~ {\rm eV},~~ \sum m_\nu > 0.1~ {\rm eV}$  for  normal and
inverted  hierarchies of neutrino masses respectively.

These predictions can be compared  with cosmological observations by
Planck \cite{Ade:2013zuv}: $ \sum m_\nu < 0.23~~{\rm
eV},~~N_{eff}=3.30 \pm 0.27$\footnote{These numbers are based on the
analysis of CMB with inclusion of Baryon Acoustic Oscillations (BAO).
However, the  results depend on the dataset used, such as  WMAP
polarisation, high resolution CMB from ACT and SPT.}.  This is 
consistent with the SM, but cannot rule out new  physics, which can
potentially change  $N_{eff}$. The possible candidates that can lead
to deviations from the SM prediction contain light $\sim 1$ eV sterile
neutrinos (see, e.g. \cite{Hamann:2010bk}),   quintessence
\cite{Wetterich:1987fm,Ratra:1987rm}, dilaton \cite{Gorbunov:2013dqa},
relic gravitational waves \cite{Giovannini:2002qw},  or new particle
decays during Big Bang Nucleosynthesis \cite{Ichikawa:2007jv}.

\section{Conclusions}
\label{concl}
During the last year a remarkable progress has been achieved both in
the field of particle physics and cosmology. The LHC experiments have
lead to the triumph of the Standard Model of particle physics, with
discovery of 126 GeV Higgs boson. The data from the Planck satellite
provided more arguments in favour of validity of the Standard
$\Lambda$CDM  cosmology.

Still, a number of fundamental questions remains unanswered, and I
present below my personal wish-list for new experiments and precision
measurements. 

In the domain of particle physics, to understand whether our vacuum is
stable or metastable and whether there is a necessity for any
intermediate energy scale between the Fermi and Planck scales
(together with the closely related issues of the Higgs inflation and
asymptotic safety of the Standard Model) we should know the mass of
the Higgs boson, the top Yukawa coupling and strong coupling constant
with highest possible accuracy. This is one of the arguments if favour
of the future  $e^+e^-$ collider - top quark factory, where a
precision measurement of the top quark mass is possible. 

The experiments that can shed light on the origin of baryon asymmetry
of the Universe  and the nature of Dark Matter are of great importance.
Quite paradoxically, the largely unexplored up to now domain of
energies where the new physics can be hidden is related to physics
{\em below} the Fermi scale. It may be very well that the new
particles giving rise to BAU and DM  are light and very weakly
interacting (as in the $\nu$MSM, briefly discussed here). Then the
search for them is not possible at the LHC but would require new
dedicated experiments \cite{Bonivento:2013jag} at the intensity and
precision frontier of high energy physics.

There is a number cosmological and astrophysical experiments, which
can elucidate  the structure of the underlying theory. In particular, 
to test the Higgs inflation, and to distinguish it from other models,
such as  $R^2$, we need to have measurements of the spectral index
$n_s$ of scalar perturbations at the level of $10^{-3}$. In addition,
the  determination of tensor-to-scalar ratio  should be done down to
values $r\simeq 0.003$ and that of the running of the spectral index 
$dn_s/d log k$ down to $5\times10^{-4}$. These measurements are
possible, presumably, at COrE \cite{COrE}, PRISM \cite{PRISM} and  SKA
\cite{SKA}. 

To check the predictions of different dynamical Dark Energy models
such as, for example, the Higgs-dilaton \cite{GarciaBellido:2011de}, 
the equation of state of dark energy (the pressure to energy density
ratio)  should be known with accuracy of $1\%$, achievable at the
Euclid mission \cite{Euclid}.

To test the infrared sector of the theory, associated with massless or
very light particles, we should know neutrino masses from cosmology
with accuracy $\sum m_\nu\simeq 0.05$ eV, to reach the lower bound
coming from particle physics. The accuracy in determination of
effective number of massless degrees of freedom below $\sim 0.04$, to
check the SM prediction  $3.046$. According to \cite{Audren:2012vy,
Basse:2013zua}, this should be  possible with the use of a combination
of  Planck and Euclid data.

The search for dark matter particles should not only be restrained to
WIMPS or axions. The dark matter may have a different nature. An
example includes the sterile neutrino with the mass in the keV region,
the radiative decays of which can be observed  with the help of high
resolution X-ray telescopes (see, e.g. \cite{Boyarsky:2012rt}).

I thank F. Bezrukov, A. Boyarsky, D. Gorbunov and O. Ruchayskiy for
helpful comments and discussions. This work was supported by the Swiss
National Science Foundation.

\end{document}